# Strangeness, Charm and Beauty in Quark Matter: SQM 2007 Experimental Overview

Federico Antinori[a,b]

[a] Istituto Nazionale di Fisica Nucleare, Padova, Italy
[b] CERN, Geneva, Switzerland

**Abstract**

This paper aims at providing an experimental overview of the Strangeness in Quark Matter 2007 Conference.

## 1. Introduction

I think it is appropriate to start this experimental summary of the SQM 2007 conference with a bit of a disclaimer: this selection is certainly going to be incomplete and biased: there is just too much material – and too little space – for a complete, detailed summary. I had to make choices, and I decided first of all to concentrate the scope on truly flavour results. Some of the results presented at this conference covered a broader spectrum than just flavour. I think that this is good: although this community has a particular interest in flavour measurements, in the end it is one single system whose properties we are studying, and I find it just appropriate that these conferences dedicate some time to hear the latest from the other fronts. On the other hand, when it came to making a selection for this summary, restricting to flavour results was the most logical choice. Within the flavour-related results, then – needless to say, I guess – I have made a personal selection. I have not tried to be ecumenical. I did try to do my best to share what I have taken home from this beautiful conference.

This paper is organized as follows: the next section is dedicated to strangeness results. Section 3 discusses the main news we heard on the production of quarkonia and – the only, slight, detour from flavour – dilepton production. Section 4, the longest, is devoted to open heavy flavour. I then conclude in section 5.

## 2. Strangeness

Strangeness is one of the "historic" observables in ultrarelativistic nucleus-nucleus collisions. In elementary collisions, strangeness production is suppressed relative to the production of light flavours. For a large system, this suppression is expected to be

removed – i.e. strangeness production is expected to be enhanced – provided the system allows for a thermodynamical description in which strangeness conservation can be treated over a large volume and does not need to be enforced locally at each collision, or, in other words, provided the correlation volume is also large, as expected if partons are freed in a Quark-Gluon Plasma (QGP) [1,2]. This is non-trivial: if, for instance, nucleus-nucleus collisions were just a simple superposition of independent nucleon-nucleon collisions, no strangeness enhancement would be expected, no matter how large the system size. In such a hypothetical case, there would be no mechanism of long-range communication within the system, and the correlation volume would still be that of each individual nucleon-nucleon collision: the system would be large, but it wouldn't "know". In other words a large system size, by itself, is not sufficient to get enhancement; a mechanism of long-range communication, sufficiently "fast" on the timescale of the collision (a few fm/c) – as would be provided by colour deconfinement – is also needed.

A pattern of strangeness enhancement increasing with the particle's strangeness content was indeed observed in Pb-Pb collisions at the SPS [3]. The most recent data on hyperon enhancements from the SPS were presented at this conference by both NA49 [4] and NA57 [5]. Reasonable agreement is found between the two experiments. Hadronic transport models are unable to reproduce these data (see for instance [6]), while they have an obvious explanation within a deconfined scenario.

The centrality dependence of the enhancements is also an important piece of information in order to constrain the models. The "canonical suppression model" [7] – for instance – predicts a fast saturation of the enhancements with the event centrality, at odds with the data. The energy dependence is also of primary interest in this respect. For the most central collisions, the enhancements do show an energy dependence, decreasing as the energy is increased from low SPS energy ($\sqrt{s_{NN}}$ = 8.73 GeV) to top SPS energy ($\sqrt{s_{NN}}$ = 17.3 GeV) [5] and to RHIC ($\sqrt{s_{NN}}$ = 200 GeV) [8]; this dependence, however, is much weaker than had been predicted [9]. I think it is fair to say that neither the centrality dependence nor the energy dependence of hyperon enhancements is yet well understood.

New results from STAR where shown at this conference on the hyperon enhancements in Cu-Cu collisions at RHIC (figure 1) [10] – found to be very similar to those measured in Au-Au collisions at corresponding values of the number of participants – and on the φ enhancement in Au-Au and Cu-Cu collisons [11] – found to be similar in the two systems and intermediate in value between those of the Λ and Ξ hyperons.

Besides the indications for statistical recombination at hadronization from the strangeness enhancement pattern already observed at the SPS, new evidence, at the kinetic level, has emerged from RHIC, with valence quark recombination counting rules seemingly controlling the different behaviour for mesons and baryons of the transverse momentum ($p_T$) dependence of the high-$p_T$ suppression and of the azimuthal asymmetry (see, for example, [12] and references therein). Despite the rather intuitive picture and the numerical successes, the physics of recombination is still somewhat unclear. For instance, it has been suggested that $\Omega^-$ production in heavy-ion collisions should be dominated by recombination of three thermal strange quarks, with combinations involving one ore more shower strange quarks strongly suppressed [13]. One would therefore expect $\Omega^-$ production to be strongly suppressed in jets relative to thermal production. Yet, STAR data show a same-side peak (a typical indication of "jetty" behaviour) associated with $\Omega^-$ produced in central Au-Au collisions [14].

The suppression of particle production at high transverse momentum, as measured for instance by the nuclear modification factor Rcp, is one of the trademark results of the RHIC programme. Taken together with other RHIC results, the effect finds a natural explanation in terms of parton energy loss in a hot and dense strongly interacting medium formed in the heavy-ion collision [15-18]. At this conference, results on the nuclear modification factors measured at the SPS were shown for $\pi^\pm$, p and $K^\pm$ [4] and for $K^0_s$ and $\Lambda$ [5] and compared with the RHIC results. The production of particles at large transverse momentum at the SPS is found to be suppressed relative to the expected amount of Cronin effect, but there is no absolute evidence for suppression, since the Rcp values stay above or around 1 at high $p_T$. The relative Rcp pattern for $K^0_s$ and $\Lambda$, however, is found to be remarkably similar between the SPS and RHIC (figure 2) [10].

New results on $K^0_s$ production were presented at this conference by NA45 [19]. The results were obtained with an original approach, aimed at maximizing the acceptance at low transverse momenta. The analysis makes use of all possible combinations of a positive and a negative track in each event, without relying on the separation between the $K^0_s$ decay vertex and the primary vertex. This results in a signal/background ratio of the order of $10^{-3}$. The $K^0_s$ peak is then extracted by subtracting the mixed-event background. Results on kaon production at the SPS are therefore now available from NA45, NA49 and NA57. A comparison is shown in figure 3. The agreement is reasonable, considering the very different experimental techniques and systematics of the various measurements.

Thermal models, as usual, are doing well. New examples from RHIC were shown at this conference [10]. A remarkable regularity from AGS to RHIC energies is apparent in the freeze-out systematics in the ($\mu_B$, T) plane [24], where $\mu_B$ is the baryon chemical potential and T the temperature extracted from the fit. The exact physical meaning of the single values of the parameters extracted from $4\pi$ fits, though, is not straightforward, due to the large variations of the baryon density, and hence of $\mu_B$, as a function of rapidity [24, 25]. Anomalies appear in the values of the thermodynamical parameters extracted from the study of particle production at the low end of the SPS energy range [26], associated with the peculiar energy dependence of strangeness production observed by NA49 [27]. New expeditions to this region are being organized. At lower energy still, FOPI again finds that a thermal fit does a very good job in describing particle production in Al+Al collisions at an incident energy of 1.9 GeV per nucleon [28]. The ($\mu_B$, T) point extracted from the fit, though, seems to lay above the curve describing the chemical freeze-out systematics at other energies and the chemical freeze-out temperature T extracted from the fit ($\sim$ 70 MeV) is lower than the kinetic temperature extracted from the particle spectra ($\sim$ 90 MeV), in contrast again with the established systematics (see [28] for details). These results call for further study.

Let's now turn to resonances. The study of the production of short-lived states could provide information on the late phases of the collision, between chemical and kinetic freeze-out, during which interactions with the medium might result in modifications in the resonances' yields, masses and widths [29]. Some of the yield ratios involving resonances seem indeed to deviate from their thermal equilibrium values, and if this were due to medium effects, such as regeneration or rescattering, it could be possible, for instance, to extract information on the system's timescales, by studying how the yields of resonances of different lifetimes are differently modified. STAR reports an interesting measurement of the $v_2$ azimuthal asymmetry coefficient for the K* particle (figure 4) [29].

The behaviour of $v_2$ versus $p_T$ is found to be well described within the by-now familiar, recombination-inspired constituent quark counting rules, assuming a number of constituents $n = 2$, as expected for K* formed by valence quark coalescence at the hadronization stage, and at odds with the behaviour expected for $n = 4$, as for K* formed in the hadronic phase by coalescence of a K and a π, with $n = 2$ constituent quarks each. STAR also reports some shift in the value of the K* mass from the PDG [30] value at low $p_T$ [31], potentially an indication of in-medium effects. The shift, though, seems to be present for all collision systems, including d-Au and p-p, and is of the order (some MeV) of the typical shifts in the mass spectra of $K^0_s$ and other particles in many experiments, due to instrumental effects.

New results are reported by NA60 [32] on the production of ϕ at SPS, where a long-standing discrepancy exists between NA49 and NA50 concerning both the yield and the slope of the $p_T$ spectra [33, 34]. The $p_T$ slopes reported by NA60 for various centrality bins are in good agreement with those of NA49, and incompatible with those of NA50. The extraction of the absolute yields in NA60 is still in progress. NA45 had also previously reported good agreement with NA49 on ϕ production, concerning both the shape of the transverse momentum distribution and the absolute yields [35].

**3. Quarkonia (and dileptons)**

While many models predicted that the J/ψ suppression would increase when going from SPS to RHIC, due to larger energy density at RHIC [36-38], the J/ψ suppression measured at RHIC has turned out to be comparable to that observed at the SPS [39]. A better agreement between predictions and RHIC data is obtained if some mechanism of J/ψ regeneration is introduced [40-44]. In this case, the similarity of the observed suppression at RHIC and SPS would be the result of a (numerically coincidental) cancellation of the extra suppression at RHIC by an increase in the abundance of charm, and hence an enhanced probability for $c\bar{c}$ recombination at hadronization. It has also been suggested, based on recent lattice calculations, that the situation may be such that, both at the SPS and at RHIC, the ψ' and $χ_c$ states, which feed down to J/ψ, could be completely suppressed, while all the directly produced J/ψ would survive [45], thus providing an explanation for the similarity of the observed suppression at the two energies. At the LHC we should actually be able to tell the two scenarios apart: if $c\bar{c}$ recombination is important at RHIC, it will be dominant at the LHC, due to the large increase in the $c\bar{c}$ production cross section. One would therefore expect J/ψ suppression to be dramatically reduced, or even to turn to an enhancement [46].

An intriguing result was obtained by PHENIX on the J/ψ suppression at high rapidity: the suppression in the interval $1.2 < |y| < 2.2$ is found to be larger than in the central region ($|y| < 0.35$) [47]. If one imagines the suppression to be connected to the energy density, one would expect lower densities, and therefore the suppression to become weaker, not stronger, as one moves away from mid-rapidity. A lead to follow up.

A nice piece of news for the quarkonia fans: STAR has observed, in pp collisions, a ϒ signal [8]. The extracted yield fits reasonably well with the cross-section systematics from other experiments.

NA60 have presented results on the study of the excess at intermediate masses (between the ϕ and the J/ψ peaks) in the invariant mass spectrum of dimuons at the SPS [32]. The excess reported by NA50 in Pb-Pb collisions is confirmed also in Cu-Cu collisions. The excess increases with the event centrality and decreases with the dimuon transverse momentum. The distribution of the weighted offset of dimuons in the mass region under study is fitted to a combination of the contributions from prompt dimuons (with an offset distribution taken to be the same as for the measured J/$\psi$ sample) and charm dimuons (for which the offset distribution is taken from a PYTHIA simulation of $c\bar{c}$ events). When both normalizations are left free, the resulting fit requires an excess of prompt dimuons by over a factor two with respect to the expected Drell-Yan contribution. The fitted amount of charm requires a value for the $c\bar{c}$ cross-section compatible with the one extrapolated from the NA50 p-A measurements, but about twice as large as expected from the interpolation of data obtained at other energies [48]. The measurement is obviously rather sensitive to the assumptions made on the transverse momentum distribution and $\Delta\phi$ distribution of $c\bar{c}$ pairs in the simulation.

PHENIX has presented data on the mass spectra of dielectrons measured in p-p and Au-Au collisions at RHIC [49]. When the spectra for the two systems are compared, an enhancement at low masses (below the $\rho/\omega$ peak) is seen in the Au-Au sample. There is also some indication for a suppression in the mass region intermediate between the $\phi$ and the J/$\psi$ (charm suppression?). Dilepton spectra do not seem to have exhausted their stock of surprises yet.

**4. Open Heavy Flavour**

Heavy flavours are produced in the initial parton-parton interaction, in the early stages of the collision. Their production can be calculated to a reasonable degree of precision within perturbative Quantum Chromo Dynamics (pQCD) and they offer the possibility to explore the medium with probes of known mass and colour charge. The energy loss by gluon radiation, for instance, is expected to be parton-specific (stronger for gluons than for quarks due to the larger colour charge of gluons [50]) and flavour-specific (stronger for lighter than for heavier quarks, due to the dead cone effect [51, 52]). Since no extra production of charm and beauty is expected at the hadronization stage, the relative rates of heavy flavour hadrons with different light quark content should be particularly sensitive to recombination effects (the $D^+_s/ D^+$ ratio, for instance, is of obvious interest in this context). In addition, the measurement of open heavy flavour production is of an essential practical interest for quarkonium physics as well: the production of open $Q\bar{Q}$ pairs is the natural reference for quarkonium studies and B meson decays are a non-negligible source of non-prompt J/$\psi$ in high energy collisions.

Since the study of heavy flavour production in heavy-ion collisions is still a very recent development, I think it is appropriate to begin this section by recalling some of the details of the experimental techniques involved.

Most of the information we have is from the RHIC "non-photonic" electron spectra [53, 54]. In STAR, electrons are identified using the measurement of the particle's energy loss in the experiment's Time Projection Chamber (TPC), combined with a time-of-flight measurement at low $p_T$ and information from the electromagnetic calorimeter at high $p_T$. PHENIX, instead, identifies electrons by combining information from the Ring Imaging CHerenkov (RICH) detector and from the candidate's energy/momentum ratio, with the energy measured in the experiment's electromagnetic calorimeter.

The next step is to reject non-heavy flavour electrons, most of which are of "photonic" origin, i.e. emerging from photon conversions or Dalitz decays of $\pi^0$ and $\eta$ ("internal" conversions). In STAR, these are dealt with by performing an invariant mass analysis of all the combinations between a candidate electron and a candidate positron in each event and removing candidates involved in low invariant mass combinations (and hence compatible with photon conversions). In PHENIX, the photonic contributions are evaluated by two independent methods (one relying on simulation and the other based on information from a special data sample collected with an additional converter of precisely known thickness inserted in the apparatus) and subtracted.

STAR also has the possibility of studying the production of $D^0$ particles via the extraction of a $D^0 \to K^-\pi^+$ signal (see for instance [55]). The $D^0$ candidates are formed by combining tracks from the TPC, with kaons and pions identified by their dE/dx. All combinations are taken, without separation between the primary vertex and the $D^0$ decay vertex. This results in a large combinatiorial background. The subtraction of the mixed-event background still leaves a substantial residual contamination, which is simply fitted and subtracted.

Another approach to the study of semileptonic decays in STAR avoids the problem of photon conversions by relying on the measurement of the production of muons, which, due to the μ-π mass difference, can be separated from pions by dE/dx in the TPC and by time-of-flight measurements at low $p_T$ (less than 100 MeV or so) [56]. The distribution of the distance of closest approach (dca) to the primary vertex of the resulting sample of muon candidates is fitted with a superposition of the shapes expected by simulation for muons from charm decays (peaking at lower dca values) and muons from pion/kaon decays (peaking farther away from the primary vertex). The method of course relies rather heavily on the accuracy of the simulation, but provides a useful independent cross-check of the other measurements.

Let's now turn to the results, starting from proton-proton collisions. PHENIX reports an excess by about 70% with respect to the central value of the pQCD Fixed-Order plus Next-To-Leading Log (FONLL) calculation [57] in the production of non-photonic electrons in proton-proton collisions at RHIC [58, 59]. The excess is rather independent of transverse momentum, with the upper value of the theoretical uncertainty band still compatible with the data. A similar result has also been reported at Tevatron by CDF [60]. The production of non-photonic electrons in p-p collisions has also been measured in STAR [54]. There is a discrepancy between the STAR and PHENIX results, with STAR reporting a cross-section about a factor two larger than PHENIX, roughly independently of $p_T$. Moving then to heavier collision systems, both PHENIX and STAR report binary scaling for the heavy flavour cross-section [53, 61], with good internal consistency across different collision systems within each experiment. The discrepancy between the two experiments seems in fact to be pretty stable, independent of the collision system, transverse momentum and detection channel (see [62]). This may be an indication for a rather basic origin of the disagreement (things such as normalizations come to mind). Incidentally, a discrepancy by about a factor two between the two experiments seems to be there also for ϕ production [63-65]. Not surprisingly, given the substantial stability of the disagreement versus collision system and $p_T$, the nuclear modification factors measured by the two experiments are in reasonable agreement with one another [62].

Both experiments observe a substantial suppression in heavy flavour production at high $p_T$ [53, 54].

Theoretically, quenching is predicted to be different for heavy and light partons. In the BDMPS framework [66], the energy loss of a parton propagating in a QCD medium is expected to be mostly due to gluon bremsstrahlung, and is calculated to be proportional to the parton's Casimir factor, measuring its colour charge (4/3 for quarks and 3 for gluons). This leads to the prediction of a larger suppression for light hadrons, expected to originate predominantly from gluon jets, than for the heavy flavoured hadrons, produced in the fragmentation of quark jets. Within the quark sector, then, heavy quarks are predicted to lose less energy than light quarks, due to the so-called dead cone effect, whereby a massive parton cannot lose energy by radiating gluons below a minimum angle, determined by the parton's mass/energy ratio [67, 68]. Due to the combination of the two above effects, heavy-flavoured particles are therefore expected to be less "quenched" than their light-flavoured counterparts. Experimentally, though, this does not seem to be the case: non-photonic electrons spectra show essentially as much high-$p_T$ suppression as charged hadrons. Ways of reconciling theory and data are being explored. It has been suggested that, in contrast to the BDMPS scenario [66], collisional energy loss may actually be important, and even dominant for low momentum partons with respect to radiative energy loss, and should therefore be taken into account [69]. The inclusion of collisional energy loss does indeed reduce the disagreement with non-photonic electron high-$p_T$ suppression data [54, 70]. Collisional losses, however, are also expected to increase the quenching for light flavours (see for instance [70]), for which the data are instead reasonably well described without invoking any additional mechanism. It should therefore be verified whether one can get a fully consistent description of all the quenching data, within the radiation-plus-collision scenario. The possibility of a reduced contribution to non-photonic electron spectra from beauty production is also being considered, although beauty decays are in principle expected to be dominant above a few GeV of $p_T$ [57]. Since, given their lower mass, c quarks are expected to lose more energy than their b counterparts due to the dead-cone effect discussed above, if the production of non-photonic electrons at RHIC were dominated by charm decays throughout the accessible $p_T$ range, it would indeed still be possible to recover a reasonable agreement with the data [71], as can be seen, for instance, in figure 5 [54]. Experimentally, this is a rather complicated matter, since the RHIC experiments are not currently equipped with microvertex detectors, which, besides providing an experimental cross-check as to the heavy flavour origin of the non-photonic electron spectra, would allow the reconstruction of secondary decay topologies, and therefore to measure separately the b and c components.

An interesting step towards b/c separation has been taken in STAR, with the study of the angular correlations between electrons and charged tracks in pp collisions [72]. Semileptonic decays of heavy flavoured hadrons introduce angular correlations between electrons and charged tracks. Such correlations are less pronounced for beauty than for charm decays, due to the larger mass, and hence larger transverse kick, for B decay electrons. The distribution of the azimuthal distance between non-photonic electrons and charged tracks in pp events is fitted with a sum of two curves with the shapes expected from simulation for charm and beauty events. The b/c ratio is then adjusted to fit the data. The method is obviously very dependent on the accuracy of the particle species mix and momentum distributions of charm and beauty hadrons in the simulation, but it is certainly very interesting to see the result. At face value, one gets a b/c ratio compatible with the FONLL prediction, although somewhat on the low side of it. Incidentally, since in STAR charm production is found to be enhanced by a factor about five relative to the FONLL prediction, the above result would imply a substantial discrepancy with respect to FONLL also for beauty production.

Another approach being explored in pp collisions in STAR is the study of the azimuthal correlation of electrons $D^0$ candidates [55]. Events where a $c\bar{c}$ pair is produced are expected to give rise to correlations at an angle around 180°, while $b\bar{b}$ events should also introduce correlations at small angle, from cases where an electron and a $D^0$ are produced in the decay of the same B hadron. Both types of correlation are visible in the preliminary data shown by STAR at this conference. Let's stay tuned.

Whatever happened to the expected difference between heavy and light quark energy loss due to the colour charge effect? Given that heavy flavour hadrons are produced in the fragmentation of quark jets, shouldn't we be seeing at least a difference in heavy vs light flavour quenching due to the factor 9/4 in the colour charge between gluons and quarks? Actually at RHIC a 6 - 8 GeV pion produced at central rapidity probes the region of x around $10^{-1}$, where the quark contribution is still large. At the LHC, the same pion will probe an x around $10^{-3}$. The quark contribution for light flavours at such transverse momenta should therefore really be negligible at the LHC, and, if the current picture is correct, the quark/gluon difference in the quenching should finally appear. We shall see.

At this conference, we were also reminded of another important – and sobering – experimental caveat: due to the different values of the semileptonic branching ratio for different charmed particles, the global semileptonic branching ratio for charm hadrons depends on the relative abundances of the different charm species. Variations in the $\Lambda_c$/D or $D_s$/D ratios when going from p-p to central Au-Au (due, say, to recombination effects) would result in a ($p_T$-dependent) modification of the global semielectronic branching ratio for charm and would therefore influence the $R_{AA}$ of non-photonic electrons even in the absence of any other effect [73, 74]. As long as we do not have control on the distribution of charm among the various species, therefore, such uncertainties will unavoidably affect the systematics.

This is just one more indication of the necessity of reconstructing secondary decay vertices. The full reconstruction of D decays will provide us with a qualitatively different, much more direct measurement of charm production with respect to the present situation, where we essentially rely on background-subtracted spectra of electrons, without any vertex separation. Vertex detectors will allow us to separate charm from beauty production, to study the relative abundances of the different hadron species in the charm sector, and to measure directly the momentum distributions of charm hadrons, without having to rely on the correlation with the momentum of the leptons emitted in the decay. The ALICE experiment (see [75, 76]), currently in the final phases of construction at the LHC, is equipped with a Silicon microvertex detector, and is expected to have very good performance for heavy flavour physics [76]. The ATLAS and CMS experiments, while designed for the study of pp collisions, will also be used for the study of heavy-ion collisions [77-79]. Both experiments are also equipped with microvertex detectors. The PHENIX and STAR experiments at RHIC are also preparing vertex detector upgrades

[80]. We shall very soon have access to powerful new tools for the study of heavy flavour production in ultrarelativistic nucleus-nucleus collisions.

## 4. Conclusions

The study of flavour production in heavy ion collisions is going stronger than ever, with more and more information on the properties of the system formed in heavy-ion collisions coming from flavour-related measurements, as was clearly visible in the breadth, depth and quality of the talks presented at this conference.

What's more, we are at the dawn of a new era, with two quantum leaps just behind the corner: the arrival of vertex detectors, with the ability to separate heavy flavour decays from the primary vertex and to study fully reconstructed heavy flavour hadrons, and the jump to the LHC energy, ushering the era of high rates and truly hard probes (imagine what we could do once we have heavy-flavour-tagged, fully reconstructed jets!).

The prospects are bright for the future of this field and of this conference series.

## Acknowledgments

I would like to thank the organizers for the beautiful conference and for presenting me with the exciting challenge of preparing the experimental summary.

I am grateful to Christoph Blume, Giuseppe Bruno, Andrea Dainese, Julien Faivre, Johann Heuser, Christian Klein-Boesing, Christian Kuhn, Carlos Lourenço, Ginés Martinez, Andre Mischke, Tomoaki Nakamura, Emanuele Quercigh, Sylwester Radomski, Victor Riabov, Karel Šafařík, Ladislav Šándor, Ferenc Siklér and Urs Wiedemann for helping me with material, discussions and suggestions.

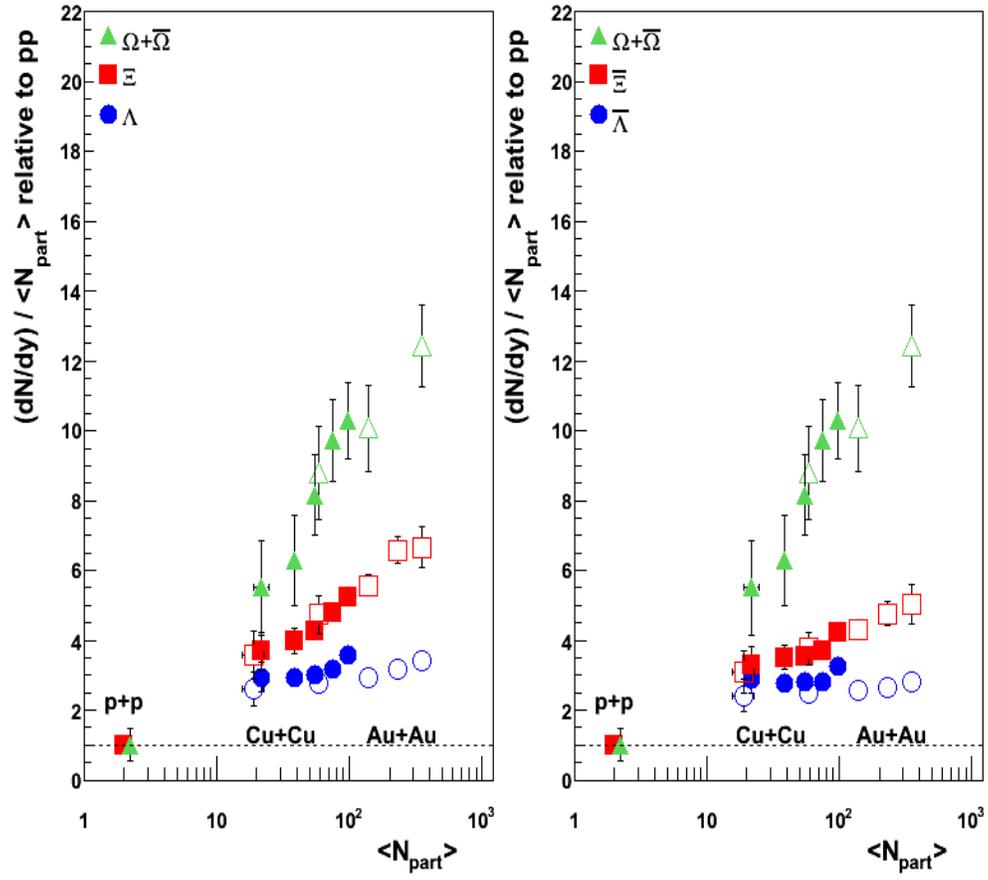

Figure 1: hyperon enhancements in Cu+Cu (closed symbols) and Au+Au (open symbols) collisions at RHIC [10]

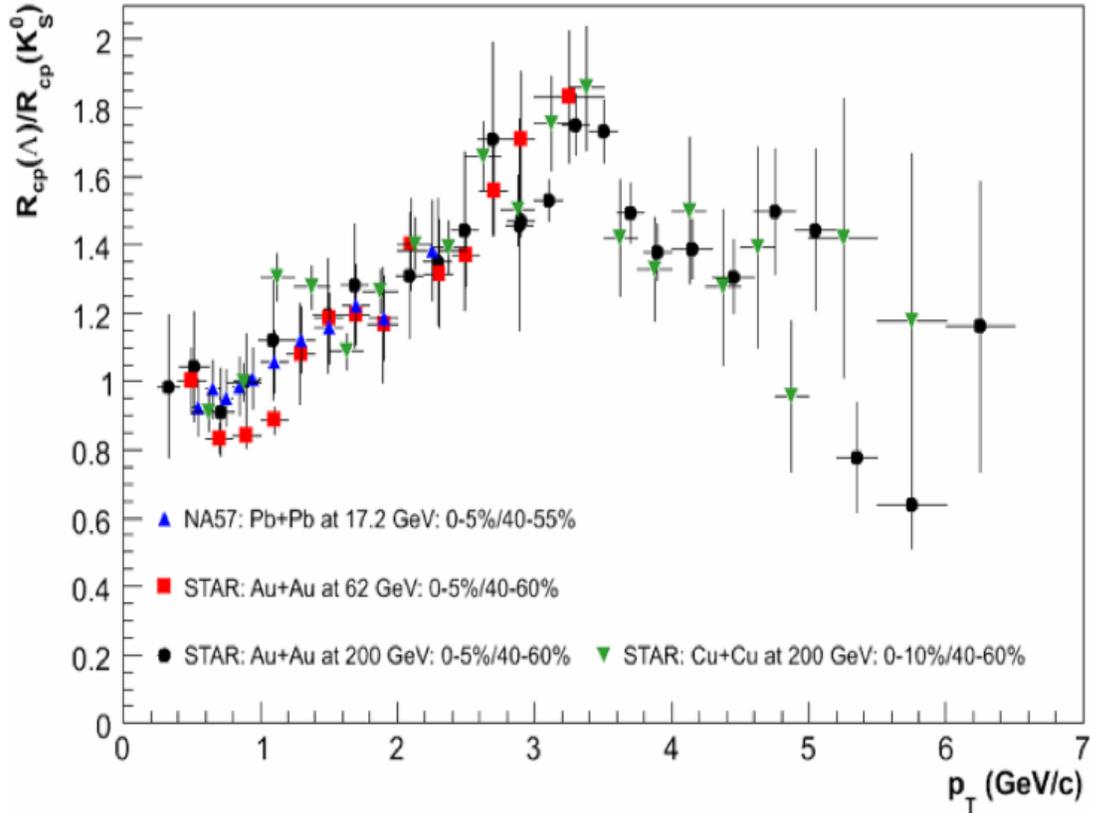

Figure 2: ratio of the nuclear modification factors for $\Lambda$ and $K^0_s$ as measured at the SPS and for different energies and colliding systems at RHIC [10]

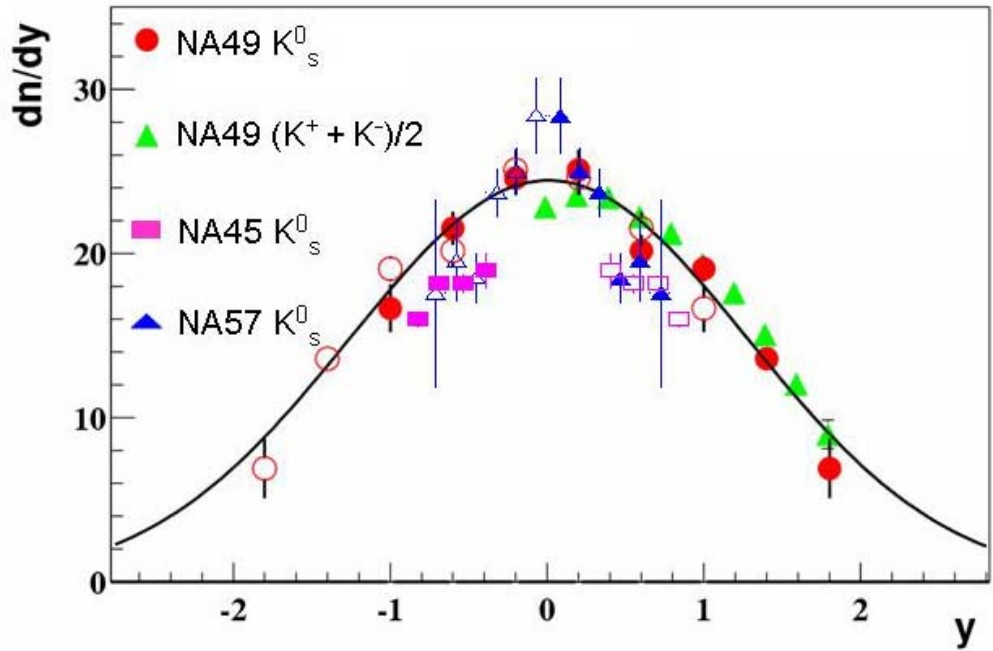

Figure 3: comparison of data on kaon dn/dy from different SPS experiments ([20], data from [19, 21-23])

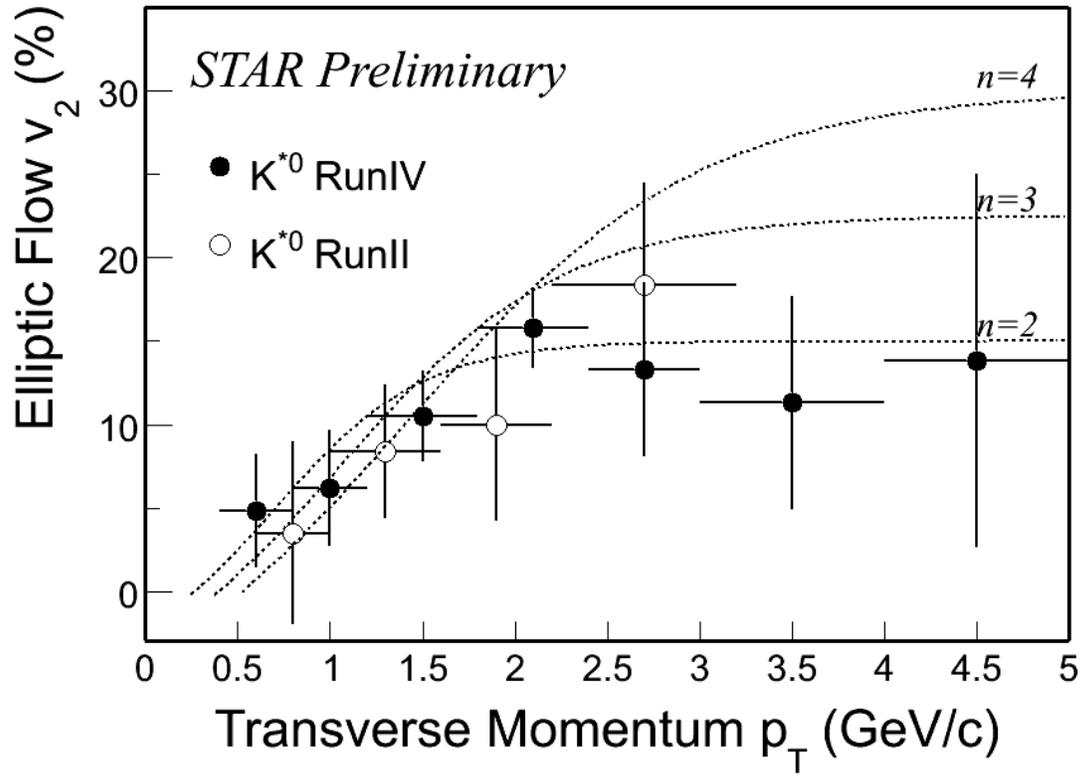

Figure 4: transverse momentum dependence of the azimuthal asymmetry coefficient $v_2$ for $K^{*0}$. The dashed curves show the expected behaviour from recombination counting rules for different values of the number of constituents $n$ [29]

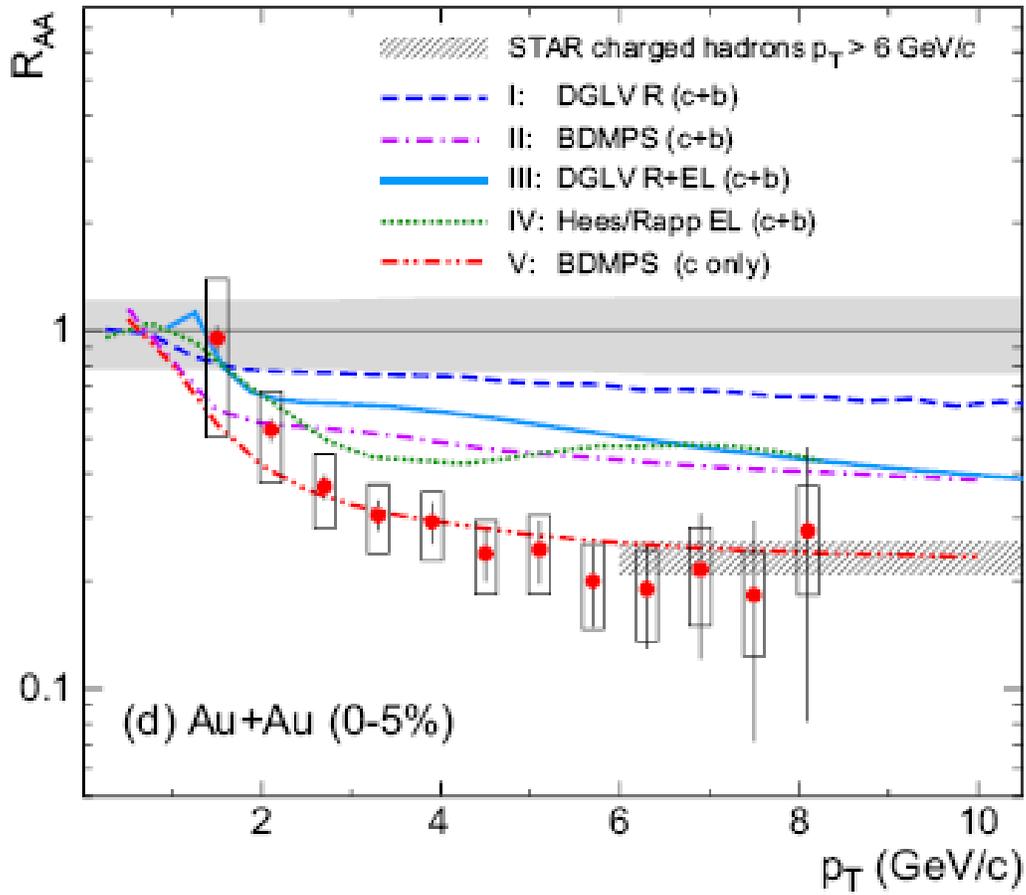

Figure 5: RAA for non photonic electrons from STAR [54], compared with theoretical curves. In particular: Curve II shows the BDMPS prediction including c and b components [71], while the BDMPS prediction for c only is given by curve V [71].